# Design research, eHealth, and the convergence revolution

Pannunzio, Valeria\*; Kleinsmann, Maaike; Snelders, Dirk

Delft University of Technology, Delft, the Netherlands
\* v.pannunzio@tudelft.nl

The Quadruple Aim is a framework which prioritizes four 'aims', or dimensions of performance, for innovating in the healthcare domain, respectively: 1) enhancing the individual experience of care; 2) improving the work life of health care clinicians and staff; 3) improving the health of populations; and 4) reducing the per capita cost of care. In this contribution, recent literature providing examples of design research in the eHealth domain is reviewed to answer the research question: 'in which measure has design research contributed to each of the 'four aims' of eHealth innovation in the past five years?'. The results of the review are presented and employed to draw three main observations: 1) design researchers in eHealth seem to be largely focused on improving experiences of care, either patients' or health professionals'; 2) design researchers' contribution on reducing per capita costs of care appears to be less pronounced, which is outlined as a point for improvement; and 3) in a considerable amount of reviewed contributions, design researchers appear to be contributing to multiple 'aims' at once. In this sub-group of reviewed contributions, several disciplinary areas and types of stakeholders interact and integrate through design research activities.
The latter observation leads to a reflection on the strategic role of design research in the contexts of the convergence revolution and of the non-communicable disease crisis. Implications of this reflection for design researchers are recognized in the opportunity and timeliness to develop eHealth-specific ways to orchestrate design integration. A direction for further research in this sense is identified in the use of sensory and self-monitored data as a boundary object for eHealth innovation. The prospective value of this direction is finally exemplified through the case of blood pressure.

## 1  Introduction

### 1.1  Design research in eHealth

eHealth is defined as the 'the application of information and communications technologies (ICT) across the whole range of functions that affect health' (Silber, 2003). In this paper, we set out to explore recent literature reporting design research case studies in the eHealth field, with the aim of understanding the specific effects and influences afforded by design researchers in this domain. Specifically, we collected eHealth-related examples of design research in the two acceptations of what Horvath (2007) calls *Research in Design Context (RiDC)* and *Design-Inclusive Research (DIR),* while discarding examples of *Practice-Based Research (PBR)* (ibid.).

For instance, literature describing usability tests conducted on eHealth proposition for design purposes (RiDC) was included in this review, as well as literature providing accounts of eHealth-relevant findings obtained through design activities (DIR). Conversely, literature providing heuristics and guidelines for designing eHealth propositions (PBR) was excluded from the review. This was chosen because, in this stage, our interest lies in understanding effects and influences afforded by design research in eHealth, rather than in exploring the practical aspects of designing in the eHealth domain.

## 1.2 The Quadruple Aim framework

The framework here employed to distinguish between *kinds* of influences afforded by design research in eHealth is the 'Quadruple Aim', a widely adopted prioritization of four dimensions of performance for improving the quality of healthcare systems. The four dimensions are depicted in Figure 1.

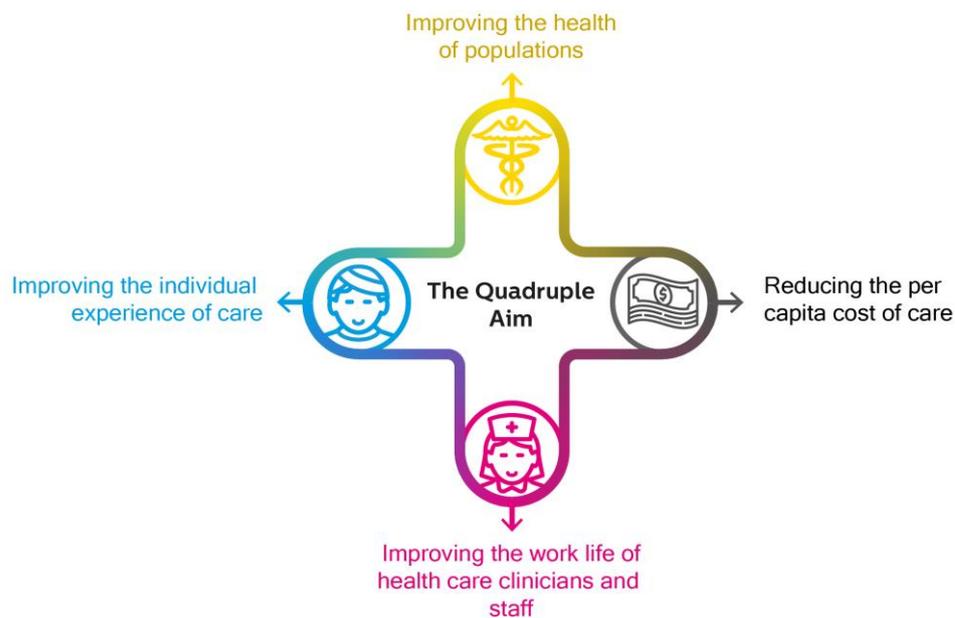

*Figure 1. The Quadruple Aim framework (authors' own illustration).*

The framework arises from a recognition of the intrinsic interconnectedness of the four dimensions; specifically, improving the health of populations is seen as the primary measure of performance of any part of a healthcare system, and the other three dimensions are seen as secondary measures of performance, all instrumental in the achievement of the former (Sikka, Morath, & Leape, 2015). The framework is recognized as pertinent to the eHealth domain, and has been successfully employed to assess the impact of specific eHealth innovations (Liddy & Keely, 2018). Exploring the impact of design research processes on each of these four aims is intended to be an exercise which is deemed useful both to stimulate awareness and self-reflection for design research practitioners working in this domain, and to serve transdisciplinary eHealth teams, whose members might not always know which kind of value to expect from design research expertise.

The overall research question is formulated as; in which measure has design research contributed to each of the 'four aims' of eHealth innovation in the past five years?

## 2  Methods

The literature review is executed as follows;

- Step 1. Advanced searches were performed in three academic databases, namely IEEExplore Digital Library, Elsevier Science Direct, and ACM Digital Library, using a combination of keyword developed iteratively and reported in Table 1. This set of databases was chosen in reason of its coverage of multiple 'flavours' of eHealth literature, including the medical-oriented (represented by sources such as the International Journal of Medical Informatics) and the computer science-oriented ones (represented by sources such as or Pervasive and Mobile Computing). We focused on the past five years, so the search is performed on papers from 2014 onwards. This step resulted in a first selection of 785 papers.

*Table 1. Keywords used for database searching.*

| eHealth keywords | "e-health"; "eHealth";"digital health"; "health IT" |
|---|---|
| Design Research keywords | "design research";"user centered design";  "patient centered design"; "user experience"; "user research" |

- Step 2. We scanned the abstracts of the papers found in Step 1 in order to exclude contributions irrelevant to the research question.
- Step 3. The remaining contributions were read in full text and excluded if deemed by the authors that; a) the contribution does not describe a single case study; b) the contribution content is not to be regarded as an example of design research as defined in the introduction; c) none of the four goals of the Quadruple Aim framework are explicitly mentioned as an objective or as an achievement of the design intervention described in the contribution. Additionally, during the review and selection process, it was decided to exclude d) four contributions that were deemed to be only indirectly health-related (e.g. describing design projects aimed at designing a website accessible for user with disabilities), thus unfit to be scrutinized through the Quadruple Aim framework; and e) one contribution that was not fully written in English. An overview of the number of contributions excluded during each of these steps is provided in Figure 2.

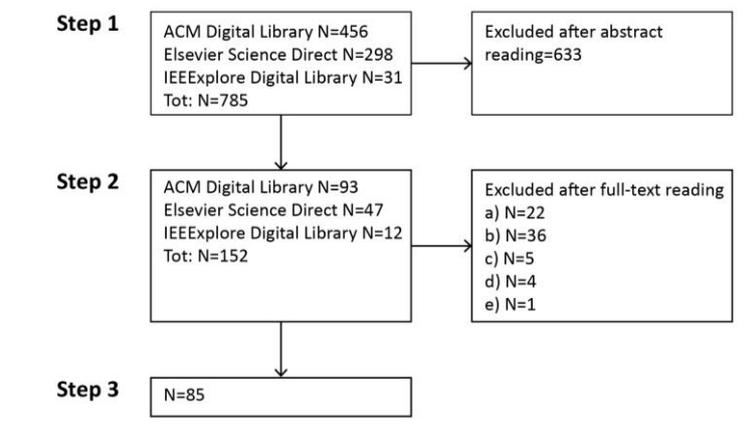

*Figure 2. Diagram summarizing the first three steps of the literature review process.*

- Step 4. The remaining 85 contributions were re-read and labelled depending on their mention of design objectives or achievements pertaining to one or more of the four dimensions of the Quadruple Aim framework. For instance, contributions mentioning 'improved patient satisfaction' as a goal or a result of an intervention supported by design research were labelled as pertaining to the first dimension, 'enhancing the individual experience of care'; contributions mentioning 'quantifiable improvements on the Healthy Eating Index' were labelled as pertaining to third dimension, 'improving the health of populations', and so on. Each contributions could be labelled on multiple dimensions.

## 3 Results

An overview of the raw results from the literature research is offered in Figure 3.

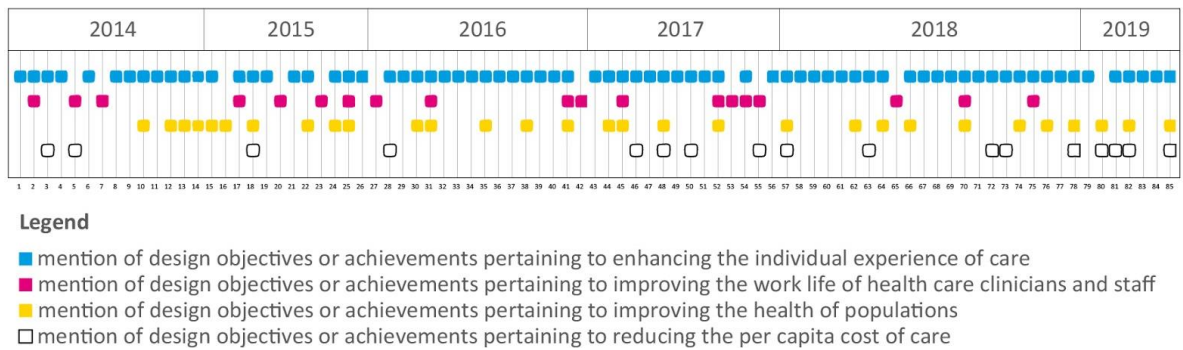

*Figure 3. Each one of the 85 contributions, represented as a vertical line, is labeled depending on the categories of aim mentioned as a design objective or achievement.*

To better understand the interconnectedness between the four aims, the contributions were also grouped based on their labels (Figure 4).

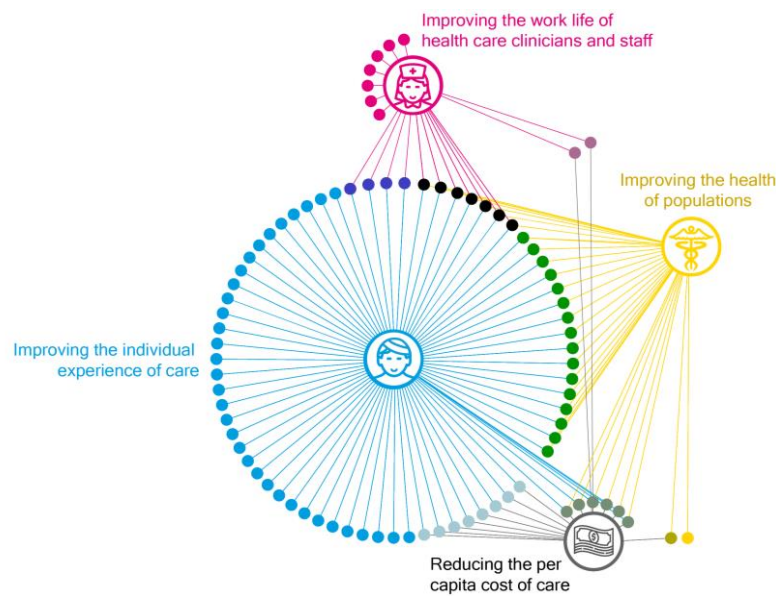

*Figure 4. The 85 contributions, represented as a coloured dot, are linked to the relevant aims (each depicted by a coloured icon). Dots are coloured depending on their connections; contributions only mentioning improved individual experiences of care are coloured in blue, contributions only mentioning improved health of populations are coloured in yellow, contributions mentioning both are coloured in green, and so on. Authors' own illustration.*

# 4 Discussion

Figures 3 and 4 collectively provide an impression of the kinds of benefits authors mention as related to design research in eHealth, and can lead us to a number of reflections.

## 4.1 Design research as the enabler for improved eHealth experience

Observing Figure 3, we can see how almost all reviewed contributions contain explicit mentions of the aim of enhancing the individual experience of care. The few contributions not mentioning individual experience-related goals tend to be the ones that do mention improving the work life of health care clinicians and staff *instead* (as it is the case of contribution 5, 7, 20, 23, 27, 42, 53, 55, and 65). This effect of mutual exclusion is easily explained by looking at target users; some design research processes are simply situated in contexts in which healthcare staff members are intended to be the primary users of the innovation - see e.g. the case of Zarabzadeh et al (2016), who investigates the utility of an electronic Clinical Prediction Rules (eCPR) amongst physicians. Altogether, thus, a first element that stands out from the overview is the strong focus on user *experiences* in eHealth design research literature, whether patients' or healthcare staff's.

The second most-often mentioned benefit of design research activities in the eHealth domain relates to improving the health of populations. A considerable number of contributions indicate specific improvements in term of health outcomes reached through or supported by design research activities. Crucially, these contributions almost never mention improved health outcomes as the *only* kind of benefit afforded, but rather as one that is coupled with improved experiences of care. As we can see in Figure 5, green dots (contributions mentioning both improved individual experiences of care and improved health of populations) appear to be fairly common cases in the reviewed literature. Reading through these contributions, two main kinds of mechanisms emerge in the way design research connects improvements in individual experiences and improvements in care outcomes, respectively;

1. Design research activities that set out to promote individual experiences of care for existing eHealth propositions, and end up impacting on care outcomes in the process - see for instance the case reported by Bakker, Kazantzis, Rickwood & Rickard (2018), in which the effort to develop an easy-to-use and engaging application resulted in a eHealth innovation which was, then, deemed to deserve its own Randomized Clinical Trial.
2. Design research activities that set out to promote improved individual experiences *so that* new, disruptive eHealth innovations that are already known to present health benefits become 'good enough' to be used - see for instance the case reported by Calvillo-Arbizu et al., (2019), in which a user-centered design process is followed so to 'maximize user acceptance' of an otherwise defined eHealth innovation.

The existence of both mechanisms, which we could refer to as 'experience-driven' and 'experience-enabled' care improvements, represent firstly a confirmation of the insights that lie at the very basis of the Quadruple Aim framework, such as the realization that care outcomes and experiences of care are inextricably linked; and secondly, a confirmation of the value of doing design research in the eHealth domain as a way to generate both 'pull' and 'push' care innovations. This last consideration aligns to theoretical models of design impact in healthcare systems, in which a distinction is drawn between a) design approaches in which design-generated knowledge is employed to develop a product or service, and b) design approaches in which design-generated knowledge is employed to develop a product or service *and* to trigger new health research (Pannunzio, in press).

**4.2     Cost-awareness in eHealth design research: a point for improvement**

Yet, the presented results should not only provide reassuring confirmations to design researchers working in the eHealth domain, but also raise puzzling concerns. The relative disinterest of design research practitioners in reducing per capita costs of care through eHealth innovations shown in Figure 3, if indeed representative of the larger eHealth scene, would be particularly alarming. In the current context of aging population and increasing prevalence of resource-intensive chronic diseases (Bloom et al, 2012), lack of cost-awareness would represent a regrettable missed opportunity for design researchers working in the eHealth domain - a field born with the very promise of providing cost-effective solutions to modern health challenges (see e.g. Stroetmann, Jones, Dobrev & Stroetmann, 2006). If eHealth becomes no more than another way to develop expensive care propositions, no matter how desirable and impactful in terms of care outcomes, the unsustainable economic burden put on modern health systems by current epidemiological trends stands few chances to be relieved.

**4.3     Multiple-aim and multi-disciplinary design research: an ally for the convergence revolution**

A conclusive reflection can be conducted on the overall landscape of design research in eHealth and its disciplinary implications. eHealth is, in fact, a realm described as inherently interdisciplinary (Pagliari, 2007; Van Velsen, Wentzel, & Van Gemert-Pijnen, 2013), in which diverse branches of knowledge - medicine, engineering, computer science, social sciences - come together and occasionally collide. Example of such 'collisions' are, for instance, the newborn fields of;

- infodemiology - described as ' the science of distribution and determinants of information in an electronic medium, specifically the Internet, or in a population, with the ultimate aim to inform public health and public policy' (Eysenbach, 2009), and
- synthetic biology, the field of study in which engineers and biologists come together to re-engineer living organisms (Khalil & Collins, 2010).

In the eHealth realm, design research can form different kinds of disciplinary bonds, some of which can be observed in the results of the literature research. Specifically, observing the overview provided in Figure 4, and keeping in mind our precedent observations, we can operate a division of the overall eHealth design research scene into three main 'zones' of transdisciplinary integration (Figure 5);

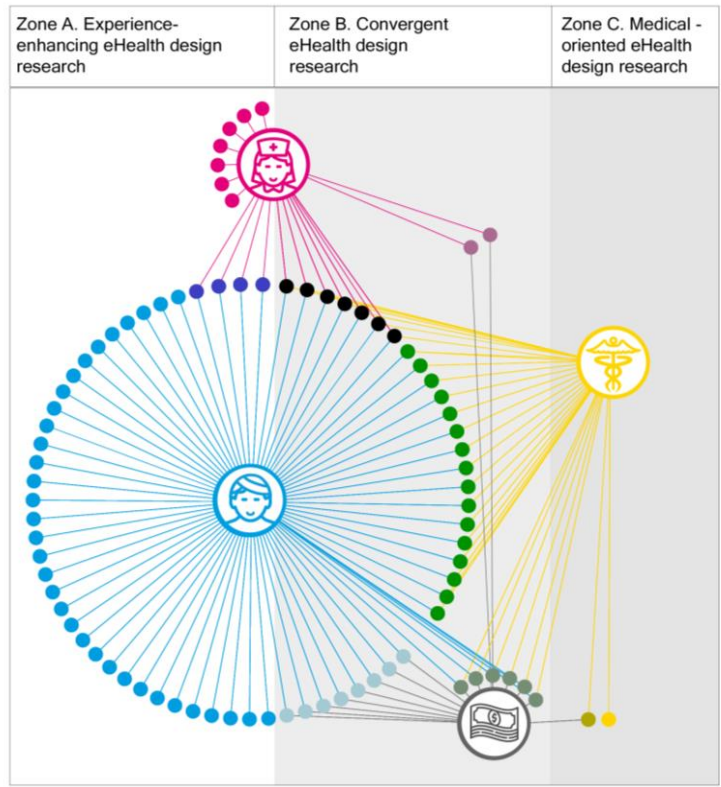

*Figure 5. eHealth design research map, distinguishing three 'zones' of design research in the eHealth field. Authors' own illustration.*

- Zone 1, in which eHealth challenges are tackled mainly from the user experience perspective (either patients', healthcare staff's, or both). Here, space for relevant transdisciplinary integration is identified between design research and disciplines such as Human Factors Engineering and Psychology.
- Zone 2, in which eHealth challenges are tackled in an integrated fashion. Here, space for relevant transdisciplinary integration is identified between design research and disparate disciplines, such as Health Service Research, Business Strategy, Industrial and System Engineering, and Computer Science.
- Zone 3, in which eHealth challenges are tackled mainly from the health outcomes perspective. Here, space for relevant transdisciplinary integration is identified between design research and medical disciplines.

This last snapshot of the eHealth design research scene is, possibly, the most intriguing one to look at to surmise upcoming developments in the field. The existence in literature - and outside of it - of a number of examples in which design research is used to address diverse sets of care goals *at the same time* through the development of eHealth innovations, as we see happening in Zone 2, allows us to recognize the strategic relevance of design research in a future perspective of convergence.

Convergence, according to Sharp, Hockfield & Jacks (2016), is the 'integration of historically distinct disciplines and technologies into a unified whole that creates fundamentally new opportunities for life science and medical practice'. Some scholars have written of the 'convergence revolution' as a third revolution in the health sciences after the discovery of DNA and the sequencing of the human genome (Ranganathan, 2017).

The Convergence Revolution, which is described as ongoing, is however not enabled by one breakthrough discovery, but rather arises from an integrated approach to the pursuit of health innovation.

### 4.4 Exploring the need for integrated approaches to health innovation: the non-communicable disease crisis

The value and timeliness of adopting an *integrated* approach on health innovation can be best understood by looking at large-scale healthcare modern challenges such as the non-communicable disease crisis. On a global level, non-infectious, or non-communicable diseases (NCDs) have been on the rise for decades, largely as a result of historical successes in the fight against infections (Figure 6).

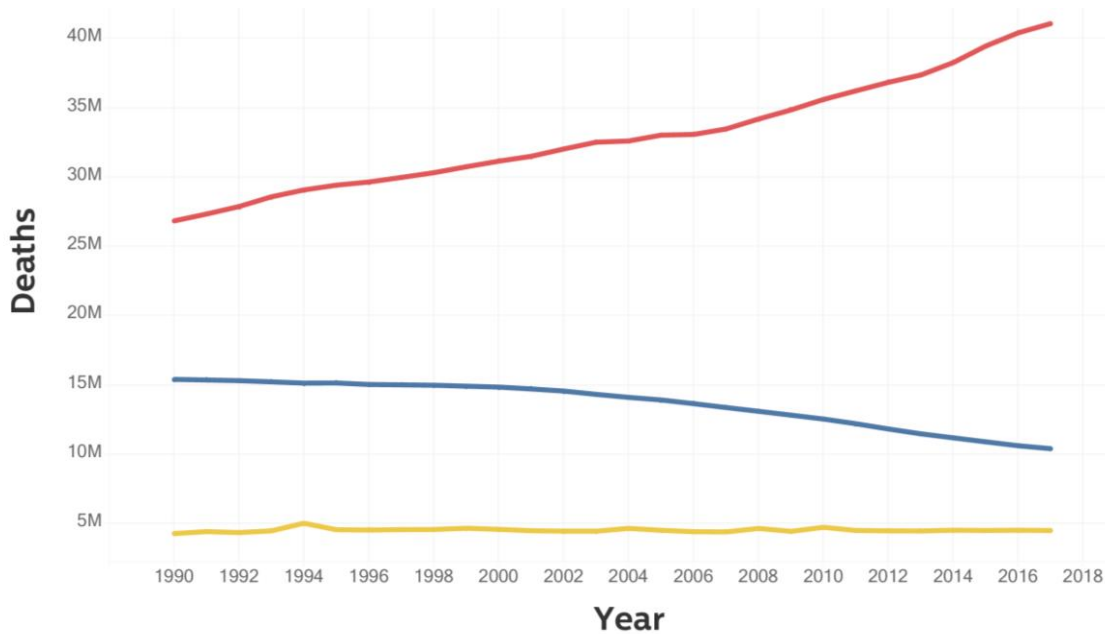

*Figure 6. Causes of death globally from 1990 to 2017 (latest data available). Authors' own illustration. Data source; Institute for Health Metrics and Evaluation, 2019a.*

Among these NCDs, four disease categories stand out (Figure 7.); cardiovascular disease, cancer (and other neoplasms), diabetes, and chronic respiratory diseases.

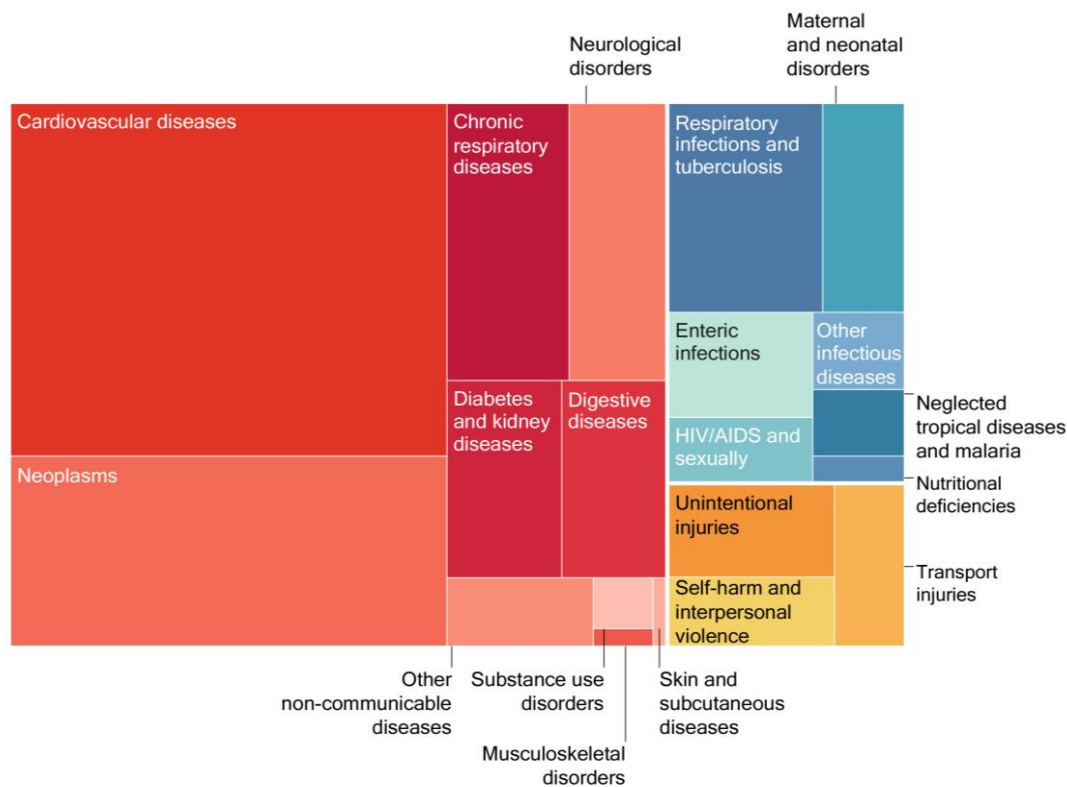

*Figure 7. Causes of death globally in 2017 (latest data available) per disease category. Non-communicale diseases are depicted in shades of red; communicable, maternal neonatal and nutritional diseases are depicted in shades of blue; and injuries are depicted in shades of yellow. Authors' own illustration. Data source: Institute for Health Metrics and Evaluation, 2019b.*

The economic impact of the non-communicable disease crisis is staggering; it is forecasted that the total cost of these conditions between 2012 and 2022 will exceed 30 trillion US dollars, damaging global GDP growth and 'pushing millions on people below the poverty line' (Bloom et al., 2012). The rise of NCDs also determines an increased demand for social- and health-care which contributes to the global shortage of health workforce, projected to result in a potential deficit of 18 million health workers by 2030 (World Health Organization, 2016). In 2011, the United Nations acknowledged in a resolution adopted by the General Assembly that 'the global burden and threat of non-communicable diseases constitutes one of the major challenges for development in the twenty first century' (United Nations, 2011). The same resolution states that prevention 'must be the cornerstone of the global response' to NCDs. Prevention is not only recognized as 'the only approach that will ensure future generations are not at risk of premature death' (Beaglehole et al., 2011a), but also as the strategy with the greatest potential to alleviate NCDs unbearable costs and workforce toll - since 'once an NCD develops, the burden on health systems (...) is substantial' (Beaglehole et al., 2011b). Following the UN high-level meeting in 2011, the World Health Assembly set a target of a 25 percent relative reduction in overall mortality from the four deadliest NCDs by 2025 (World Health Organization, 2013). However, the latest progress monitor, covering data up until 2017, reported that 'progress has been insufficient and highly uneven' (World Health Organization, 2017).

The insufficient progress should not surprise; preventing NCDs on a population level is a challenge that presents unprecedented difficulties for health systems. NCDs, in fact, tend to develop as results of a complex interplay of concurrent causes, or risk factors. As we can observe in Figure 8., typical risk factors for NCDs include dietary, physical activity or smoking behaviours.

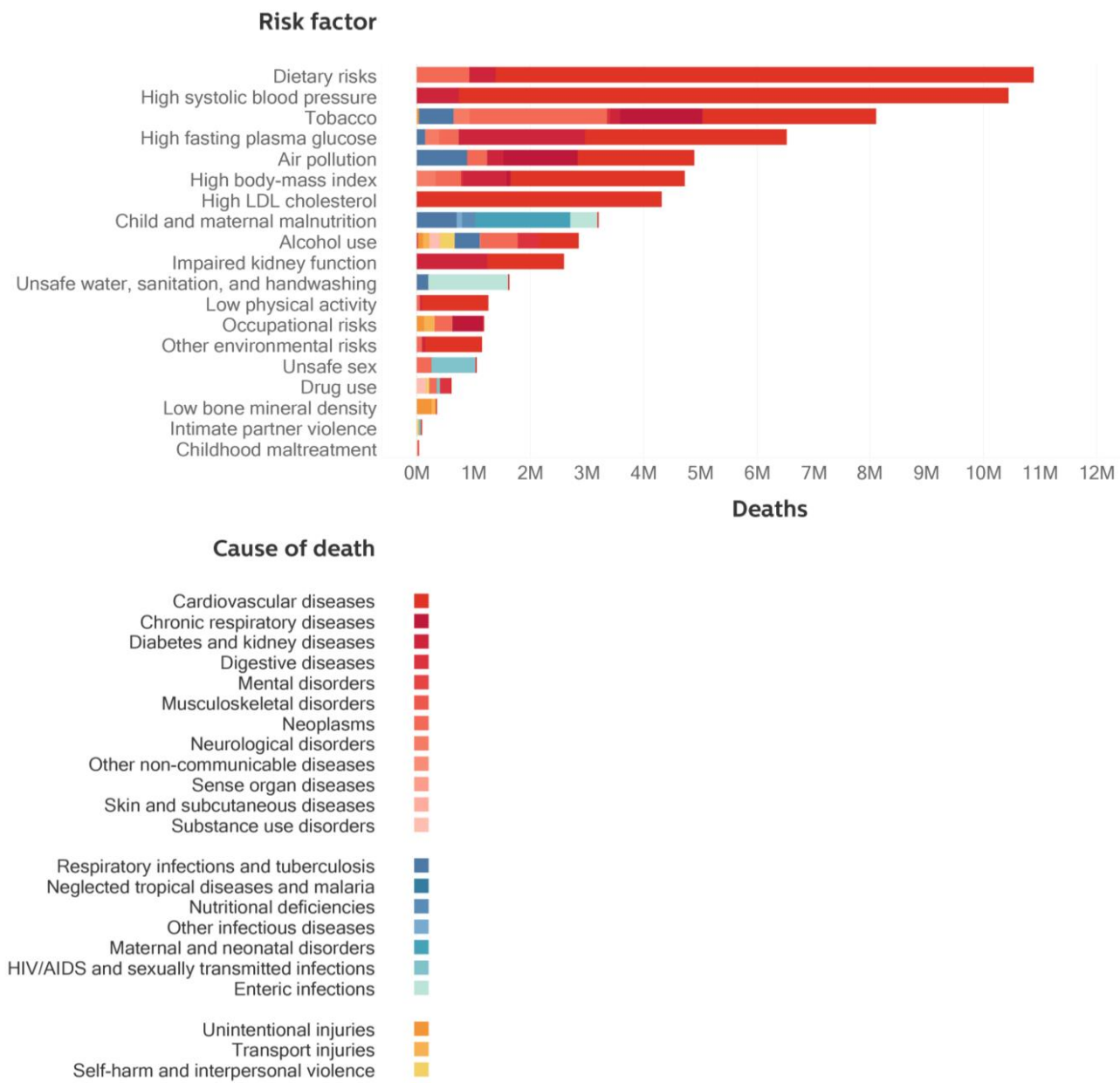

*Figure 8. Risk factors linked to causes of death globally for 2017 (latest data available). Authors' own illustration. Data source: Institute for Health Metrics and Evaluation, 2019c.*

Preventing NCDs through reduction of these kinds of risk factors means, in practice, getting individuals who do not have a disease to adopt healthier behaviours - for consistent amounts of time. Healthcare-specific capabilities find themselves ill-prepared to cope with a similar task; after all, both clinical disciplines and material systems of health practice evolved in rather different conditions and responding to the needs of *sick* individuals. Widespread, direct healthcare interventions towards *non-sick* individuals can be extremely efficient from a clinical point of view, but clash into problems that are beyond healthcare's disciplinary reach, and are more closely linked to historical, cultural, political and contextual factors. Worrisome, for instance, is the example of vaccination campaigns, one of the greatest achievements of public health, and yet among the interventions that generate the most long-lived controversies within subsets of the public (Dubé, Vivion & MacDonald, 2015).

To cope with such 'externalized' health challenges, effective integration - of capabilities, contexts, and functions –is crucial. One good example of this principle can be found in the case of tobacco consumption: the reduction of smoking habits in a number of high- and middle-income countries is regarded as one the biggest successes so far in the control of NCDs risk factors (Ezzati & Riboli, 2013). Such result is deemed to have been driven by measures such as restrictive taxation, smoke-free policies in public spaces, warning labels, and bans on advertising promotion and sponsorship (Gravely et al., 2017). The implementation of these measures is described as a successful integration of achievements from different disciplines - production of clear scientific evidence regarding the harms tobacco consumption, execution of careful cost-effectiveness estimations, and innovative developments in legislation (Shibuya, 2003).

In other areas of NCD risk, examples of effective integration are yet to be found; the lack of effective measures for improving diet and exercise, in particular, led some to define overweight, obesity, and high blood glucose as the 'wild cards' of global NCD risks, and to call for 'bold, creative policies that address harmful alcohol consumption, improve diet, and increase physical activity' (Ezzati & Riboli, 2013). Others advocate the need for an "interdisciplinary social and behavioral approach, including the cultural aspects of nutrition" (Bousquet et al., 2011). Convergence, with its promise of integrating 'historically distinct disciplines and technologies', presents itself as an ideal approach for exploring 'what it means to be well, to function at the peak of our physical and mental capabilities, as well as to prevent or deal with illness' (Sharp et al., 2016).

## 5 Directions for further research

### 5.1 Design integration through digital data

The ability to integrate and connect different contexts and specialized disciplines is identified as a core design capability in design literature (Kleinsmann & Valkenburg, 2008). Buchanan, for instance, (1992) elaborates on design as an integrative discipline, which connects knowledge from the arts and sciences in ways that are appropriate to the problems and purposes at hand. Dorst (1997) provides a detailed account of integration as a design activity, which he identifies as 'a reasoning process building up a network of decisions (part of the design problem or the design solution) while taking account of different contexts (distinct ways of looking at the problem or solution)'. Still recently, the integrative power of design and its specific value in the health domain has been examined by Romm & Vink, (2018), who elaborate on the 'in-betweenness' of service design practitioners working in healthcare.

This integrative power appears to be especially necessary in a context of increasing convergence, in which health innovation is expected to arise from stakeholders afferent to different disciplines - each one with their own 'ways of looking'. We observe this design ability in action in the results of the present literature review, and specifically in the examples that populate 'Zone 2' (Figure 5.).

Doing design research in convergent eHealth scenarios becomes, thus, not only a matter collecting and producing knowledge, but also a matter of reconciling different types of knowledge and orchestrating their contribution in the design process. Orchestrating service co-creation for the purpose of planning and carrying out knowledge integration activities was, indeed, recently recognized as a strategic design ability for integrated care innovation (Durón, Simonse & Kleinsmann 2019).

A designerly way in which this orchestration can be managed is through the use of boundary objects, or artefacts that are 'both plastic enough to adapt to local needs and constraints of the several parties employing them, yet robust enough to maintain a common identity across sites' (Star, 1989). Carlile (2002) identifies three characteristics of 'effective' boundary objects in new product development, being;

1. (The boundary object) establishes a shared syntax or language
2. (The boundary object) provides a concrete means for individuals to specify and learn about their differences and dependencies across a given boundary
3. (The boundary object) facilitates a process where individuals can jointly transform their knowledge (p.451).

Boundary objects can be embodied in a wide array of formats, both material and immaterial. Mortier, Haddadi, Henderson, McAuley and Crowcroft (2014) elaborate on the use of *digital data* as a boundary object in ubiquitous computing settings, in reason of the capacity of these data to be 'open to multiple interpretations and the concern of many stakeholders'. Indeed, in the eHealth domain, a unique opportunity of design-led integration is constituted by the possibility of using data (and especially sensory and patient-reported data) as a boundary object which satisfies each of the previously specified condition for effectiveness. Respectively;

1. Sensory and patient-reported data can be employed as a way to 'establish a shared syntax or language' in reason of their capacity to generate *syntheses* of complex, cross-contextual networks of meanings within eHealth design research. In one of the papers populating Zone 2. (Figure 5.), for instance, we observe how 'data-driven' medical consultations are enabled by a eHealth intervention in which clinicians can prescribe '10,000 steps a day' to patients who wish to improve their physical activity levels (Kim et al., 2017). Here, a shared syntax for doctor-patient conversation is generated by collapsing the complexity of physical activity (both a clinician-understood health metric and a patient-understood everyday life behavior) into a quantified goal that can be easily recognized by both parties.
2. Sensory and patient-reported data can 'provide concrete means for individuals to specify about their differences and dependencies across a given boundary', in reason of their capacity to surface *antitheses* in stakeholders' needs and purposes regarding a eHealth proposition. In van Kollenburg, Bogers, Rutjes, Deckers, Frens and Hummels (2018), for instance, we learn of an exploration of the value of parent-tracked baby data in interactions with healthcare professionals. Starting from parents-reported data, the design researchers could identify specific differences in how parents and health professionals envisioned a preferred care workflow (e.g. parents favoured richer data overviews while GPs preferred simpler data summaries).

3. Sensory and patient-reported data can 'facilitate a process where individuals can jointly transform their knowledge', in reason of their capacity to introduce *changes* in the knowledge bases themselves. For instance, the introduction of glucose self-monitoring devices for diabetic patients, which enabled more frequent measurements than previous technology, is described to have 'shifted the value' of the information about glucose levels, 'challenging the numerical standards for "normalcy"' (Mol & Law, 2004 as cited in Fiore-Gartland & Neff, 2015).

The use of sensory and patient-reported data as a boundary object in eHealth design research is identified as a promising strategy for design integration in a context of convergence. The entire field of medical-grade wearable sensors, specifically, which is recognized by Mertz (2016) to 'rely on' the convergence revolution, is a domain in which design researchers can effectively apply this strategy. Next, future opportunities for design research in this direction are illustrated through the case of blood pressure.

## 5.2. The blood pressure example

Unobtrusive wearable technologies for the self-monitoring of blood pressure, a crucial metric for cardiovascular health, are being developed and will become more and more common in the next decades. In January 2019, the first wristwatch able to take clinically accurate blood pressure readings was released in the American market (Omron Healthcare, 2019). According to the manufacturer's website, the product went almost immediately sold out, and to the moment in which this paper is written, aspiring customers can, at most, enrol in a waiting list.

This innovation opens new, uncharted eHealth scenarios: the early market success of the product indicates the existence of a robust demand for consumer-facing blood pressure wearable monitors, but does not help envisioning how will we, as consumers, use these wearables and the data they collect. How will this change our habits, routines and lifestyles? What opportunities will this technology afford us?

To investigate these questions, we intend to explore the use of self-monitored blood pressure data as a boundary object for the development of integrated services propositions for cardiovascular prevention. As observable in Figure 8., high blood pressure is a prominent risk factor for several NCDs, and in particular for cardiovascular diseases, the class of conditions responsible for most deaths worldwide. The development of measurable and cost-effective ways to control blood pressure in a large enough subset of the population would constitute a 'quadruple-aimed' innovation, able to;

1. improve individual experiences of care by enabling personalized, meaningful ways of managing one's own cardiovascular health
2. improve the work life of health care clinicians and staff by reducing chronic care workloads and promoting the availability of data useful for population health management
3. reduce the per capita cost of care by preventing or delaying the development of chronic, non-communicable conditions
4. Improve the cardiovascular health of populations by reducing the incidence of hypertension, especially through the adoption of healthier behaviours such as a low-sodium diet and active lifestyle, which would have preventive effects on the other main NCDs as well.

Of course, this is easier said than done; in such a challenge lie numerous, multifaceted complexities, most of which are not for design researchers to solve. Yet, it is a challenge for design researchers to surface these complexities, so that the relevant disciplines and stakeholders may use them as a way to create shared understandings, to face misalignments, or to advance themselves.

# 6  Conclusions

In this contribution, recent examples of design research in the eHealth domain were reviewed to answer the research question: 'in which measure has design research contributed to each of the 'four aims' of eHealth innovation in the past five years?'. The research results provided a snapshot of the contemporary eHealth design research scene which led the authors to three main conclusions;

1. design researchers in eHealth seem to be largely focused on improving experiences of care, either patients' or health professionals';
2. design researchers' contribution on reducing per capita costs of care appears to be less pronounced;
3. In a considerable amount of reviewed contributions, design researchers appear to be contributing to multiple 'aims' at once. In this sub-group of reviewed contributions, several disciplinary areas and types of stakeholders interact and integrate through design research activities.

From these conclusions, key contributions to the field were identified, namely; 1) a solicitation for design research working in eHealth to reserve increased attention to cost-effectiveness aspects; and 2) a call for design researchers in eHealth to embrace their strategic role in the contexts of the convergence revolution, particularly by developing new, eHealth-specific ways to orchestrate design integration. A direction for further research in this regard was identified in the use of sensory and self-monitored data as a boundary object; finally, the prospective value of this direction was exemplified through the example of blood pressure.

**About the Authors:**

**Valeria Pannunzio:** Currently a Ph.D. researcher in the CardioLab at the IDE faculty of the Delft University of Technology, Valeria holds a master's degree in Design for Interaction and previously worked at Philips Healthcare as a service designer.

**Maaike Kleinsmann:** Associate Professor in design-driven innovation at the IDE faculty of the Delft University of Technology, Maaike is also the head of the CardioLab, an initiative of the IDE faculty, De Hartstichting and Philips Design.

**Dirk Snelders:** Head of the section Methodology and Organisation of Design at the IDE faculty of the Delft University of Technology, Dirk has a background in psychology and marketing, where he has developed his research interest on the role of design in business.

**Acknowledgement:** We gratefully acknowledge the anonymous peer reviewers for the valuable support.